\documentclass[a4paper,11pt,fleqn]{article}
\usepackage[latin1]{inputenc}
\usepackage[T1]{fontenc}
\usepackage{amsfonts}
\usepackage{amssymb}
\usepackage{latexsym}
\usepackage{amsmath}
\usepackage{amsthm}
\usepackage{graphicx}

\numberwithin{equation}{section}

\newcommand{\rme}{\textrm{e}}  %
\newcommand{\rmd}{\textrm{d}}  %
\newcommand{\rmi}{\textrm{i}}  %

\title{Weak and strong expansions of the generalized $q$-deformed coherent states approximate eigenfunctions and its resolution of unity}
\author{YAHIAOUI Sid-Ahmed,
        BENTAIBA Mustapha\thanks{email address: \texttt{bentaiba@univ-blida.dz}}\\
LPTHIRM, d\'epartement de physique, facult\'e des sciences, universit\'e Sa\^ad DAHLAB-Blida 1,\\
B.P. 270 Route de Soum\^aa, 09\,000 Blida, Algeria}

\begin{document}

\maketitle

\begin{abstract}
\noindent The aim of this paper is to provide an explicit expressions for the generalized $q$-deformed harmonic oscillator coherent states obtained in terms of a weak and strong behavior expansions. We first use the weak-deformed version ($s\rightarrow0$) of $q$-boson annihilation operator to solve Barut-Girardello's eigenvalues coherent states equation for the generalized $q$-deformed harmonic oscillator. In strong behavior limit ($s\rightarrow\infty$) the previous result is resumed using the variational perturbation theory. We also describe the construction of their resolution of unity.\\[0.5pc]
\textbf{PACS numbers}: 02.20.Uw, 02.30.Gp, 02.30.Hq, 02.30.Mv\\
\textbf{Key Words}: Quantum groups, Special functions, Weak and strong behavior expansions
\end{abstract}

\section{Introduction}%

\noindent Quantum groups (QG), considered to be a generalization of the fundamental symmetry concepts of classical Lie groups, have been the subject of intensive research and several standard textbooks have been devoted to this exciting field \cite{1,2,3,4,5}. The idea to associate quantum groups to the classical groups leads to develop the concept of $q$-deformed quantum mechanics \cite{6,7,8,9,10,11,12,13} and a special attention is devoted to the $q$-deformed Weyl-Heisenberg algebra ($q$-WH) of raising and lowering operators (see for example, \cite{6,7}).\\
\indent Since QG have attracted great attention on behalf of physicists and found close applications in various fields of physics and chemistry such as quantum inverse scattering theory \cite{14}, nuclear physics \cite{15}, solvable statistical mechanics models \cite{16} and molecular physics \cite{17}. In mathematical physics, they have found to be interesting in the point of view of coherent states \cite{18,19,20,21}. In this context and in analogy with the well-known non-deformed CS, the $q$-deformed coherent states ($q$-CS) are the eigenstates of the $q$-deformed boson annihilation operator and are superpositions of $q$-deformed harmonic oscillators ($q$-HO). The $q$-deformed CS have been well studied and widely applied to mathematical physics \cite{22,23,24,25,26,27,28}.\\
\indent The $q$-deformed HO can be defined with the help of coordinate description of the creation and annihilation operators realize in a space of functions, $A_s^\dagger$ and $A_s$, introduced by Macfarlane \cite{6}:
\begin{eqnarray}
A_s=\alpha\big(\rme^{-2\rmi sx}-\rme^{-\rmi sx}\rme^{\rmi s\frac{\rmd}{\rmd x}}\big),\qquad
A_s^\dagger=\overline{\alpha}\big(\rme^{2\rmi sx}-\rme^{\rmi s\frac{\rmd}{\rmd x}}\rme^{\rmi sx}\big),\label{1.1}
\end{eqnarray}
with $\hbar=1$ and obeying the $q$-commutation ($q$-mutator) rule
\begin{eqnarray}\label{1.2}
\left[A_s,A_s^\dagger\right]_q\equiv A_s A_s^\dagger-q^2A_s^\dagger A_s=\mathbf{1}_{\mathfrak{E}},
\end{eqnarray}
where $q$ is assumed to be real and related to the parameter $s$ via $q=\rme^{-s^2}$, and is taken to lie between 0 and 1. Using \eqref{1.2} it is easy to be convinced that $\alpha\overline{\alpha}=(1-q^2)^{-1}$.\\
\indent However, in the most of cases, information about a physical system can only be obtained by means of approximation procedures, due to the fact that the underlying equations cannot be solved analytically. Many different approximation procedures are developed in order to deal with non-analytically solvable system, and the {\it perturbation theory} is by far one of the most commonly used approach. It is based upon the expansion of some physical quantity, e.g. $\lambda$, into a power series of the parameter $s$, namely,
\begin{eqnarray}\label{1.3}
  \lambda_N(s) &=& \sum_{n=0}^N \lambda_n\,s^n,
\end{eqnarray}
and the obtained results in the weak-behavior ($s\rightarrow0$) seem to converge to the exact result for low orders.\\
\indent Unfortunately, when the expansion is used to describe the strong-behavior ($s\rightarrow\infty$), the original weak-behavior expansion will cease to describe the complete system, because the divergent of that series becomes important when the expansion is driven to the higher orders. Therefore, it is necessary to {\it resum} such series in order to deal with the divergent perturbation expansions. Many resummation procedures have been proposed, amongst them we can find the variational perturbation theory (VPT). This latter allows the converting of divergent weak-behavior expansion into convergent strong-behavior expansion \cite{29}. It permits the evaluation of divergent series of the form \eqref{1.3} for all values of the parameter $s$, including the strong-behavior, and gives a strong-behavior expansion of the form
\begin{eqnarray}\label{1.4}
  \lambda_N(s) &=& \kappa^{p}\sum_{n=0}^N \lambda^{(N)}_n\,\left(\frac{s}{\kappa^r}\right)^n\Big|_{\kappa=1},
\end{eqnarray}
by introducing a scaling parameter $\kappa$, which is set to one after computation. Here the parameters $p$ and $r$ are integers and characterize completely the strong-behavior (for more details we refer the readers to \cite{30,31} and references therein). By setting {\it Kleinert's square-root} substitution \cite{29}
\begin{eqnarray}\label{1.5}
\kappa=K\sqrt{1+sk} \qquad{\textrm{with}}\qquad k=\frac{\kappa^2-K^2}{sK^2},
\end{eqnarray}
in \eqref{1.4}, the variational parameter $K$ has the strong-behavior expansion in the form
\begin{eqnarray}\label{1.6}
  K^{(N)}(s) &=& s^{1/r}\sum_{n=0}^N K^{(N)}_n s^{-2n/r},
\end{eqnarray}
and the physical quantity $\lambda(s)$ behaves like
\begin{eqnarray}\label{1.7}
  \lambda_{\rm strong}(s) &=& s^{p/r}\sum_{n=0}^N \lambda^{(N)}\left(K^{(N)}_0,K^{(N)}_1,K^{(N)}_2,\cdots,K^{(N)}_n\right) s^{-2n/r},
\end{eqnarray}
where the coefficients $K^{(N)}_n$, $n=0,1,2,\cdots,$ depend on the order $N$.\\
\indent It is the purpose of this paper to exploit $q$-WH algebra and to obtain new expansion formulas that emerge from weak- and strong-behavior limits. The focus here is on applying a weak behavior expansion of $q$-deformed boson annihilation operator \eqref{1.1} to the generalized $q$-deformed HO in order to solve the eigenvalues equation associated with the generalized $q$-deformed Barut-Girardello's coherent states (BG CS), using to this end some lemmas directly related to the resolution of the Riccati differential equation. In the sequel, VPT can be applied to resum the deduced weak-behavior expansion of the eigenvalue corresponding to BG CS equation. Finally, we discuss their resolution of unity and we prove that they admit a specific measure coinciding with the elliptic Jacobi $\vartheta_3$-function and the space of functions is considered to be the unit circle, which seems here to be suited and more appropriate.\\
\indent This paper is structured as follows. In section 2, we introduce a weak version of $q$-deformed boson annihilation operator whose its characteristics enables us to describe the eigenfunctions for the generalized $q$-deformed CS in terms of the Riccati equation. Section 3 is dedicated to some features integrability lemmas of the Riccati equation. Section 4 is devoted to the construction process of the weak-behavior expansion ($s\rightarrow0$) for the generalized $q$-deformed HO CS, followed by applying VPT in order to resum the obtained series to deal with the divergent perturbation expansions ($s\rightarrow\infty$). In section 5, we prove that the constructed weak $q$-deformed HO CS admit a unity resolution relation and expressed through a positive-definite weight function coinciding with the elliptic Jacobi-function. Finally, last section contains the conclusion.

\section{Weak deformed approximation and Riccati equation}%

We begin by specify the coordinate representation for the creation and annihilation operators related to our object of study. By generalized $q$-deformed HO, we mean that $x$ is just replaced by the function $\beta(x)=x+b(x)$ in \eqref{1.1} keeping the operator $\frac{\rmd}{\rmd x}$ unchanged. It is obvious that we recover the standard HO when the function $b(x)=0$. Then we suggest to rewrite \eqref{1.1} in the form
\begin{eqnarray}
A_s\rightarrow\mathcal{A}_s&=&\alpha\left(\rme^{-2\rmi s\,\beta(x)}-\rme^{-\rmi s\,\beta(x)}\rme^{\rmi s\frac{\rmd}{\rmd x}}\right),\label{2.1}\\
A_s^\dagger\rightarrow\mathcal{A}_s^\dagger&=&\overline{\alpha}\left(\rme^{2\rmi s\,\beta(x)}-\rme^{\rmi s\frac{\rmd}{\rmd x}}\rme^{\rmi s\,\beta(x)}\right),\label{2.2}
\end{eqnarray}
where $\beta(x)$ is some function to be determined. This determination comes from the restriction that the ladder operators in \eqref{2.1} and \eqref{2.2} satisfy the $q$-mutator rule \eqref{1.2} under the same constraints imposed to $q$ and $\alpha$ in the introduction, i.e. $q=\rme^{-s^2}$ and $\alpha\overline{\alpha}=(1-q^2)^{-1}$.\\
\indent Then taking into account \eqref{1.2} and applying the Campbell-Baker-Hausdorff relation, we obtain after some straightforward calculation the relation satisfying the function $b(x)$, namely
\begin{eqnarray}\label{2.3}
b(x+\rmi s)\equiv b(x),
\end{eqnarray}
which means that $b(x)$ is an arbitrary periodic function with period equal to $\rmi s$. As a consequence it is worth mentioning that a $\rmi s$-periodic function $b(x)$ of the variable $x$ can be written as a power series by defining a new variable $\frac{x}{s}$, such as
\begin{eqnarray}\label{2.4}
b(x)=\sum_{n=-\infty}^{\infty}c_n\,\exp\left(\frac{2n\pi x}{s}\right)\quad\Rightarrow\quad
b(x+\rmi s)=\sum_{n=-\infty}^{\infty}c_n\,\rme^{\frac{2n\pi(x+\rmi s)}{s}}\equiv b(x),
\end{eqnarray}
where the coefficients $c_n$ are arbitrary and $n\in\mathbb Z$. As $b(x)$ has a period of $\rmi s$, we may think of this function as the Fourier series for a function in a new variable, e.g. $t$, with period $2\pi$. This restriction may be easily relaxed by substituting $x$ by $\frac{\rmi s}{2\pi}\,t$, and we get
\begin{eqnarray}\label{2.5}
b(x)\rightarrow b(t)=\sum_{n=-\infty}^{\infty}c_n\,\rme^{\rmi nt},\qquad\mathrm{with}\qquad
c_n=\frac{1}{2\pi \rmi}\oint_{(\mathcal{C})}\frac{b(\zeta)}{\zeta^{n+1}}\,\rmd\zeta,
\end{eqnarray}
where $\zeta=\rme^{\rmi t}$. Equation \eqref{2.5} guarantees that our $2\pi$-periodic function is now constructed \textit{sectionally} on each interval $\mathcal{I}_n=[(2n-1)\pi,(2n+1)\pi]$, with the Fourier part-function $b_n(t)=\rme^{\rmi nt}$. Then it is obvious that $c_n$ are complex Fourier coefficients and due to the fact that $\zeta=\rme^{\rmi t}$, so the contour $(\mathcal{C})$ is a unit circle.\\
\indent Inspired by the property of the function $b(x)$, i.e. $\beta(x)$, discussed previously and using the fact that up to the first order we have $\alpha\simeq\frac{1}{\sqrt{2}s}+\frac{s}{2\sqrt{2}}\in\mathbb R$ for the small parameter $s$, we will, in what follows, express the $q$-deformed boson annihilation operator $\mathcal A_s$ in its weak $q$-deformed approximation scheme given by
\begin{eqnarray}\label{2.6}
\mathcal A_s\simeq\frac{s}{\sqrt{2}}\frac{\rmd^2}{\rmd x^2}
-\frac{\rmi}{\sqrt{2}}\left(1-\rmi s\beta(x)\right)\frac{\rmd}{\rmd x}
-\frac{\rmi}{\sqrt{2}}\left(1-\frac{3\rmi s}{2}\,\beta(x)\right)\beta(x),
\end{eqnarray}
in order to solve the associated $q$-deformed Barut-Girardello's HO CS eigenvalues equation, i.e.
\begin{eqnarray}\label{2.7}
\mathcal A_s|\mathbf{\Lambda},s\rangle=\lambda_s|\mathbf{\Lambda},s\rangle,
\end{eqnarray}
where in the configuration space, $\langle x|\mathbf{\Lambda},s\rangle\equiv\mathbf{\Lambda}_s(x)$, the eigenvalues equation \eqref{2.7} becomes
\begin{equation}\label{2.8}
\begin{split}
\mathcal{L}_s\mathbf{\Lambda}_s(x)&\equiv\left[\frac{\rmd^2}{\rmd x^2}-\frac{2\rmi}{s}\left(1-\rmi s\beta(x)\right)\frac{\rmd}{\rmd x}
-\frac{2\rmi}{s}\left(1-\frac{3\rmi s}{2}\,\beta(x)\right)\beta(x)\right]\mathbf{\Lambda}_s(x)\\
&=\frac{2\sqrt{2}\lambda_s}{s}\,\mathbf{\Lambda}_s(x).
\end{split}
\end{equation}
\indent As a last step of our calculations, let us look for solutions of \eqref{2.8} in the form
\begin{eqnarray}\label{2.9}
\mathbf{\Lambda}_s(x)=\xi_s(x)\exp\left[\frac{\rmi}{s}\,x+\delta\int^x\beta(\eta)\,\rmd\eta\right],
\end{eqnarray}
where $\delta$ is some constant to be determined subsequently (or to be avoided.) If one substitutes \eqref{2.9} into \eqref{2.8}, it is easy to show that the corresponding eigenvalues equation for $\xi_s(x)$ reads as
\begin{equation}\label{2.10}
\begin{split}
\frac{\rmd^2\xi_s(x)}{\rmd x^2}&+2(\delta-1)\beta(x)\frac{\rmd\xi_s(x)}{\rmd x}\\
&+\left(\frac{1-2\sqrt{2}s\lambda_s}{s^2}
-\frac{4\rmi}{s}\beta(x)+(\delta+1)(\delta-3)\beta^2(x)+\delta\beta'(x)\right)\xi_s(x)=0,
\end{split}
\end{equation}
and by means of change of function
\begin{equation}\label{2.11}
z_s(x)=-\frac{\rmd}{\rmd x}\,\ln\xi_s(x)\qquad\Rightarrow\qquad \xi_s(x)\sim\exp\left[-\int^x z_s(\eta)\,\rmd\eta\right],
\end{equation}
the differential equation \eqref{2.10} is reduced to the Riccati equation given by
\begin{equation}\label{2.12}
\begin{split}
\frac{\rmd z_s(x)}{\rmd x}=&\,z^2_s(x)-2(\delta-1)\beta(x)z_s(x)\\
&+\frac{1-2\sqrt{2}s\lambda_s}{s^2}-\frac{4\rmi}{s}\beta(x)+(\delta+1)(\delta-3)\beta^2(x)+\delta\beta'(x).
\end{split}
\end{equation}
\indent In order to obtain the solution of \eqref{2.12}, we introduce in the next section some integrability lemmas satisfying some restrictions on the coefficients of the Riccati equation.

\section{Some lemmas about Riccati equation}%

\noindent The Riccati equation
\begin{eqnarray}\label{3.1}
\frac{\rmd z(x)}{\rmd x}=p_2(x)z^2(x)+p_1(x)z(x)+p_0(x),
\end{eqnarray}
plays a significant role in many fields of applied and fundamental science and is one of the most studied first-order non-linear differential equations \cite{32}.\\
\indent It is well established that the solutions are obtained by assuming certain relations among the coefficients $p_i(x)$, $(i=0,1,2)$, of \eqref{3.1} which lead to involve some lemmas. Let us briefly review two important lemmas about the properties of the solutions of Riccati equation and solved analytically. The proofs of both lemmas are discussed and given in the cited references.
\newtheorem{Lemma}{Lemma}[section]
\begin{Lemma}[\cite{32,33}]
Let $p_2(x)=1$, $p_0(x)$ and $p_1(x)$ be polynomials. If the degree of the polynomial
$\mathcal{S}(x)=p_1^2(x)- 2p_1'(x)-4p_0(x)$ is odd, the Riccati equation can not possess a polynomial solution. If the degree of $\mathcal{S}(x)$ is even, the equation involved may possess only the following polynomial solutions:
\begin{eqnarray}\label{3.2}
z^{(\pm)}(x)=-\frac{1}{2}\left(p_1(x)\pm\left\lfloor\sqrt{\mathcal{S}(x)}\right\rfloor\right),
\end{eqnarray}
where $\lfloor\sqrt{\mathcal{S}(x)}\rfloor$ denotes an integer rational part of the expansion of $\mathcal{S}(x)$ in decreasing powers of $x$.
\end{Lemma}
\begin{Lemma}[\cite{32,34}]
The Riccati equation \eqref{3.1} is solvable by quadrature if a relationship
\begin{eqnarray}\label{3.3}
\omega_1^2\,p_2(x)+\omega_1\omega_2\,p_1(x)+\omega_2^2\,p_0(x)=0,
\end{eqnarray}
exists with constant coefficients $\omega_1$ and $\omega_2$, not simultaneously zero, and satisfying the condition $|\omega_1|+|\omega_2|>0$.
\end{Lemma}
\indent In accordance with two lemmas exposed hereinabove, we combine both of them in the next section in order to solve the Riccati equation \eqref{2.12}.

\section{Weak- and strong-expansions for the generalized $q$-deformed coherent states}%

\noindent Following Lemma~3.1, the coefficients $p_0(x)$ and $p_1(x)$ of Riccati's differential equation \eqref{2.12} are polynomials of the variable $x$ and satisfying
\begin{eqnarray}
p_0(x)&=&\frac{1-2\sqrt{2}s\lambda_s}{s^2}-\frac{4\rmi}{s}\beta(x)+(\delta+1)(\delta-3)\beta^2(x)
+\delta\beta'(x),\label{4.1}\\
p_1(x)&=&-2(\delta-1)\beta(x)\label{4.2},
\end{eqnarray}
with $p_2(x)=1$. As a consequence, these coefficients involve that the function $\beta(x)$, i.e. $b(x)$, must be a polynomial, too. Then the polynomial $\mathcal{S}(x)$ is given by
\begin{equation}\label{4.3}
\mathcal{S}(x)=16\beta^2(x)+\frac{16\rmi}{s}\beta(x)-\frac{4}{s^2}\left(1-2\sqrt{2}s\lambda_s\right)-4\beta'(x).
\end{equation}
\indent On the other hand, the Lemma~3.2 give us the possibility to solve \eqref{2.12} through quadratures. Therefore \eqref{3.3} can be express as
\begin{equation}\label{4.4}
\delta\beta'(x)+(\delta+1)(\delta-3)\beta^2(x)
-\left(\frac{4\rmi}{s}+2(\delta-1)\frac{\omega}{1+\omega^2}\right)\beta(x)-\frac{1-2\sqrt{2}s\lambda_s}{s^2}=0,
\end{equation}
where $\omega=\frac{\omega_1}{\omega_2}$ is the new parameter of quadrature ($\omega_2\neq0$) and it is considered here to be a real parameter. The next step consists in eliminating the term $\beta'(x)$ from \eqref{4.3} and \eqref{4.4}, which lead us to express the function $\mathcal{S}(x)$ as a quadratic function in $\beta(x)$
\begin{eqnarray}\label{4.5}
\mathcal{S}(x)&=&\frac{4}{\delta}(\delta-1)(\delta+3)\beta^2(x)+
\frac{4(\delta-1)}{\delta}\left(\frac{4\rmi}{s}-\frac{2\omega}{1+\omega^2}\right)\beta(x)-\frac{4(\delta-1)}{\delta}\frac{1-2\sqrt{2}s\lambda_s}{s^2}.
\end{eqnarray}
\indent Unfortunately, it seems that Lemma~3.1 as it is postulated is inconvenient for application because the polynomial $\mathcal{S}(x)$, by definition, has its integer rational part in decreasing powers of $x$ which is not necessarily the case here. Essentially all we have to do is to think about the function $\mathcal{S}(x)$ otherwise and to choose it in order to be able to apply Lemma 3.1 appropriately. In this way since $\mathcal{S}(x)$ is a polynomial and quadratic in $\beta(x)$, the expression under the square-root sign in \eqref{3.2} must be regarded as the square of a polynomial. This is possible only if the discriminant $\Delta(\mathcal{S})$ of \eqref{4.5} is equal to zero, i.e.
\begin{equation}\label{4.6}
\frac{-64(\delta-1)^2}{\delta^2s^2(1+\omega^2)^2}\left[2s(1+\omega^2)\left(2\rmi\omega+\sqrt2\lambda_s(\delta+3)(1+\omega^2)\right)
-(\delta-1)(1+\omega^2)^2-s^2\omega^2\right]=0,
\end{equation}
and hence we can extract the expression of eigenvalues $\lambda_s$ related to the parameters $\delta$ and $\omega$ and expanded up to the first-order in powers of $s$ as
\begin{equation}\label{4.7}
\lambda_{\delta,\omega}(s)=\frac{\delta-1}{2\sqrt2(\delta+3)}\frac{1}{s}-\frac{\rmi\sqrt2\omega}{(\delta+3)(1+\omega^2)}+
\frac{\omega^2}{2\sqrt2(\delta+3)(1+\omega^2)^2}\,s.
\end{equation}
\indent Now since $\mathcal{S}(x)$ is a quadratic function in $\beta(x)$ and its discriminant is equal to zero, then the expression under the square-root has a quadratic form and determine completely the polynomial $\mathcal{S}(x)$, which is given by
\begin{equation}\label{4.8}
\mathcal{S}(x)=\frac{4(\delta-1)}{\delta(\delta+3)}\left((\delta+3)\beta(x)+\frac{2\rmi}{s}-\frac{\omega}{1+\omega^2}\right)^2,
\end{equation}
where it is necessary, for the function $\mathcal S(x)$, that $\delta\neq0,1,-3.$\\
\indent Then using \eqref{3.2}, \eqref{2.11}, and \eqref{2.9}, we obtain the expressions of $z_s(x)$, $\xi_s(x)$ and the generalized $q$-deformed HO CS, $\mathbf{\Lambda}_s(x)$, respectively, up to the normalization constant
\begin{eqnarray}
z_s^{(\pm)}(x)&=&\left(\delta-\nu^{(\pm)}\right)\beta(x)-\frac{\rmi}{s}(\gamma^{(\pm)}_s-1),\label{4.9}\\
\xi_s^{(\pm)}(x)&=&\exp\left[-\left(\delta-\nu^{(\pm)}\right)\int^x\beta(\eta)\,\rmd\eta
                   +\frac{\rmi}{s}\left(\gamma^{(\pm)}_s-1\right)x\right],\label{4.10}\\
\mathbf{\Lambda}_s^{(\pm)}(x)&\sim&\exp\left[\nu^{(\pm)}\int^x\beta(\eta)\,\rmd\eta+\frac{\rmi}{s}\,\gamma^{(\pm)}_s\,x\right],\label{4.11}
\end{eqnarray}
which do not depend on the eigenvalue $\lambda_s$ and the parameters $\nu^{(\pm)}$ and $\gamma^{(\pm)}_s$ are given by
\[
\nu^{(\pm)}=1\pm\sqrt{\frac{(\delta-1)(\delta+3)}{\delta}},\qquad
\gamma^{(\pm)}_s=1\pm\sqrt{\frac{\delta-1}{\delta(\delta+3)}}\left(2+\frac{\rmi\omega}{1+\omega^2}\,s\right).
\]
\indent One can observe, due to the Lemma~3.1, that the generalized $q$-deformed HO CS in \eqref{4.11} have two possibilities and are both solutions of \eqref{2.9}. At this stage one can ask whether there are any other alternative approaches which allows us to interpret \eqref{4.11} with the eigenvalues \eqref{4.7}. Here we try to answer this question by observing that the associated eigenvalues are expressed in terms of positive and negative powers of $s$. For this reason it is helpful to use a perturbation procedure in the first case and to resum the results, using VPT, in the second case.

\subsection{Weak-behavior expansion: perturbation procedure}%

Let us expand the eigenfunctions $\mathbf{\Lambda}_s^{(\pm)}(x)$ and the eigenvalues $\lambda_s$ in terms of power series of $s$ as
\begin{equation}\label{4.12}
\mathbf{\Lambda}_s^{(\pm)}(x)=\sum_{n=0}^N \phi_{n,s}^{(\pm)}(x)s^n,\qquad\mathrm{and}\qquad
\lambda_{\rm weak}=\sum_{n=0}^N \lambda_{n,s} s^n.
\end{equation}
\indent Substituting the expansions of \eqref{4.12} into \eqref{2.8} and equating terms with like powers of $s$ leads to determine the leading weak-behavior coefficients $\phi_{n,s}^{(\pm)}(x)$ given by a series of two-equations,
\begin{gather}
  \phi_{n,s}'^{(\pm)}(x)+\frac{\rmi}{2}\,\phi_{n-1,s}''^{(\pm)}(x)-\rmi\beta(x)\phi_{n-1,s}'^{(\pm)}(x)
  -\frac{3\rmi}{2}\,\beta^2(x)\phi_{n-1,s}^{(\pm)}(x)+\beta(x)\phi_{n,s}^{(\pm)}(x)\nonumber\\
  \qquad\qquad=\rmi\sqrt2\sum_{k=0}^n\lambda_{k,s}\phi_{n-k,s}^{(\pm)}(x),\label{4.13}\\
  \phi_{n,s}''^{(\pm)}(x)-2\beta(x)\phi_{n,s}'^{(\pm)}(x)-3\beta^2(x)\phi_{n,s}^{(\pm)}(x)
  =2\sqrt2\sum_{k=1}^n\lambda_{k,s}\phi_{n-k,s}^{(\pm)}(x),\label{4.14}
\end{gather}
with the constraint on the coefficients $\phi_{n-k,s}^{(\pm)}(x)=0$ for $n-k<0$.

\subsection{Strong-behavior expansion: variational perturbation theory}%

The main purpose of VPT is the resummation of divergent series in the case of the strong-behavior expansion ($s\rightarrow\infty$) \cite{29,30,31}. Here we will apply VPT to our $q$-deformed HO. The weak-behavior expansion for the eigenvalues $\lambda(s)$ up to the first-order is given by \eqref{4.7} and the leading strong-behavior coefficients can be obtained using VPT.\\
\indent To this end, we can resum the weak-behavior series obtained in \eqref{4.7} by setting
\begin{eqnarray}\label{4.15}
  l(s) \equiv s\lambda(s) = \frac{\delta-1}{2\sqrt2(\delta+3)}-\frac{\rmi\sqrt2}{\delta+3}\,\frac{s}{\kappa^2}
                            +\frac{1}{2\sqrt2(\delta+3)}\,\left(\frac{s}{\kappa^2}\right)^2,
\end{eqnarray}
where $\kappa^2=(1+\omega^2)/\omega>1$ is chosen in order to ensure the positivity of $\omega$. The scaling-function $\kappa(\omega)$ depends only on the quadrature parameter $\omega$ and reaches its minimum at $\kappa_{\rm min}=\sqrt2$, for $\omega_{\rm min}=1$. Comparing \eqref{4.15} with \eqref{1.4} and identifying $\kappa$ as the scaling parameter, one has the strong parameters $p=0$ and $r=2$. In the sequel, inserting \eqref{1.5} and re-expanding in $s$ to the first-order, without forgetting to replace the parameter $k$ by its value in \eqref{1.5}, we have
\begin{eqnarray}\label{4.16}
  l(s) \equiv s\lambda(s) = \frac{\delta-1}{2\sqrt2(\delta+3)}+\frac{2\rmi(\kappa^2-2K^2)}{\sqrt2(\delta+3)K^4}\,s
                            +\frac{1}{2\sqrt2(\delta+3)K^4}\,s^2.
\end{eqnarray}
\indent Extremizing the modulus of \eqref{4.16}, the variational parameter $K$ can be expanded in the strong-behavior series, up to the first-order $N=1$, following \eqref{1.6}
\begin{eqnarray}\label{4.17}
  K^{(1)}(s) &=& s^{1/2}\left(K^{(1)}_0+K^{(1)}_1s^{-1}+K^{(1)}_2s^{-2}\right),
\end{eqnarray}
where the real coefficients $K^{(1)}_n,\,(n=0,1,2)$, are given in terms of $\delta$ and $\omega$ by
\[
   K^{(1)}_0=\frac{1}{(-\delta-31)^{1/4}},\quad
   K^{(1)}_1=-\frac{12(1+\omega^2)}{\omega(-\delta-31)^{3/4}},\quad{\rm and}\quad
   K^{(1)}_2=-\frac{4(\delta+13)(1+\omega^2)^2}{\omega^2(-\delta-31)^{5/4}},
\]
with $\omega\neq0$ and the condition $\delta<-31$ is taken into account in order to to ensure that all coefficients $K^{(1)}_n$ are reals. Inserting the last result and \eqref{4.17} into \eqref{4.16}, we obtain the strong-behavior series for the eigenvalue
\begin{eqnarray}\label{4.18}
  \lambda_{\rm strong}(s) = \frac{1}{s}\left(\lambda^{(1)}_0+\lambda^{(1)}_1s^{-1}+\lambda^{(1)}_2s^{-2}\right),
\end{eqnarray}
with the first-order complex coefficients $\lambda^{(1)}_n$ depend on both of $\delta$ and $\omega$, namely
\begin{eqnarray*}
  \lambda^{(1)}_0 &=& -\frac{2\sqrt2}{\delta+3}\,(4+\sqrt{-\delta-31}), \\
  \lambda^{(1)}_1 &=& \frac{\sqrt2\,(1+\omega^2)}{\omega(\delta+3)}\,\left[12\sqrt{-\delta-31}-\rmi(\delta+79)\right], \\
  \lambda^{(1)}_2 &=& \frac{4\sqrt2\,(1+\omega^2)^2}{\omega^2(\delta+3)\sqrt{-\delta-31}}\,\left[(\delta+103)\sqrt{-\delta-31}-16\rmi(\delta+40)\right].
\end{eqnarray*}
\indent Finally, the $q$-deformed CS for the generalized HO potential can be expanded in the strong-behavior series and yields
\begin{eqnarray}\label{4.19}
  \mathbf{\Lambda}_{\rm strong}^{(\pm)}(x) &=& \phi_{0,s}^{(\pm)}(x)+s^{-1}\phi_{1,s}^{(\pm)}(x)+s^{-2}\phi_{2,s}^{(\pm)}(x),
\end{eqnarray}
where $\phi_{n,s}^{(\pm)}(x),\,(n=0,1,2)$, satisfy \eqref{4.13} and \eqref{4.14}.\\
\indent Another important property to be discuss in the next section concerns the resolution of unity for a set \eqref{4.11}.

\section{Resolution of unity and its consequence}%

\noindent It is well-known that the determination of a unity resolution relation for any set of CS is indeed a difficult task, because it imposes some severe constraints on CS. In this sense we are going to prove that the generalized $q$-deformed HO CS, $\mathbf{\Lambda}^{(\pm)}_s(x)$, are endowed with a resolution of unity and expressed in terms of a certain positive-definite weight function. Our proof follows basically the formal mathematical treatment sketched in \cite{8} but differs slightly in some points.\\
\indent To demonstrate this specific identity, we first begin by defining
\begin{equation}\label{5.1}
\mathbf{1}_{\mathfrak{E}}\equiv\int_{(\mathcal{I})}\rmd\mu^{(\pm)}_s(x)\,\overline{\mathbf{\Lambda}_s^{(\pm)}}(x)\mathbf{\Lambda}_s^{(\pm)}(x),
\end{equation}
where $\rmd\mu^{(\pm)}_s(x)=\sigma^{(\pm)}_s(x)\,\rmd x$ serving as a measure in $\mathfrak{E}$ and $\sigma^{(\pm)}_s(x)$ is a real and positive-definite weight function to be determined. Here $\mathcal{I}$ stands for the domain of integration which depends closely on the generalized $q$-deformed HO CS, $\mathbf{\Lambda}_s^{(\pm)}(x)$.\\
\indent However the compactness of the physical configuration space, $\mathfrak{E}$, is unfortunately ill-defined. To solve the problem, we begin first by considering all $\mathbf{\Lambda}_s^{(\pm)}(x)$ defined in $\mathcal{I}_\infty=(-\infty,\infty)$ and we will use the change of variable of the section 2, i.e. $x=\frac{\rmi s}{2\pi}\,t$, which has an advantage to deduce the nature of the space of functions. In the other words this allows us to reduce $\mathfrak{E}$ to the unit circle which seems here to be more appropriate.\\
\indent Knowing that $\beta(x)=x+b(x)$, \eqref{4.11} can be expressed as
\begin{equation}\label{5.2}
\mathbf{\Lambda}_s^{(\pm)}(t)\sim\exp\left[-\frac{\gamma^{(\pm)}_s}{2\pi}\,t
+\frac{\nu^{(\pm)}}{2}\left(\frac{\rmi s}{2\pi}\,t\right)^2+
\frac{\rmi s}{2\pi}\,\nu^{(\pm)}\int^t b(\eta)\,\rmd\eta\right],
\end{equation}
where $\gamma^{(\pm)}_s$ and $\nu^{(\pm)}$ are given hereinabove, and $b(t)$ is $2\pi$-periodic function. Now \eqref{5.2} are well-defined over a unit circle and obeying to the relations
\begin{eqnarray}\label{5.3}
\mathbf{\Lambda}_s^{(\pm)}(t+2\pi)&\sim&\exp\left[-\gamma^{(\pm)}_s-\frac{\nu^{(\pm)}}{2}\frac{s^2}{\pi}\,t-\frac{\nu^{(\pm)}}{2}\,s^2\right]
\mathbf{\Lambda}_s^{(\pm)}(t), \nonumber \\
\mathbf{\Lambda}_s^{(\pm)}(t+4\pi)&\sim&\exp\left[-2\gamma^{(\pm)}_s-2\frac{\nu^{(\pm)}}{2}\frac{s^2}{\pi}\,t-4\frac{\nu^{(\pm)}}{2}\,s^2\right]
\mathbf{\Lambda}_s^{(\pm)}(t), \nonumber \\
\vdots &\vdots& \vdots
\end{eqnarray}
and so on. Then, by mathematical induction, we get from \eqref{5.3}
\begin{equation}\label{5.4}
\mathbf{\Lambda}_s^{(\pm)}(t+2\pi n)\sim
\exp\left[-n\gamma^{(\pm)}_s-n\frac{\nu^{(\pm)}}{2}\frac{s^2}{\pi}\,t-n^2\frac{\nu^{(\pm)}}{2}\,s^2\right]\,\mathbf{\Lambda}_s^{(\pm)}(t),
\end{equation}
which is valid for all $n\in\mathbb{Z}$. Exploiting this property, it is interesting to express \eqref{5.1} in the equivalent form
\begin{eqnarray}\label{5.5}
\mathbf{1}_{\mathfrak{E}}=\int_{-\infty}^{\infty}\rmd x\,
\overline{\mathbf{\Lambda}_s^{(\pm)}}(x)\sigma^{(\pm)}_s(x)\mathbf{\Lambda}_s^{(\pm)}(x)
\equiv\lim_{N\rightarrow\infty}\sum_{n=-N}^{+N}\int_{(2n-1)\pi}^{(2n+1)\pi}
\rmd x\,\overline{\mathbf{\Lambda}_s^{(\pm)}}(x)\sigma^{(\pm)}_s(x)\mathbf{\Lambda}_s^{(\pm)}(x),
\end{eqnarray}
where the domain of weak oscillator coordinate is considered to be covered by the infinite sum of the finite interval $\mathcal{I}_n=\left[(2n-1)\pi,(2n+1)\pi\right]$ as
\begin{equation}\label{5.6}
\mathcal{I}_{\infty}=\bigcup_{n=-\infty}^{\infty}\mathcal{I}_n,
\end{equation}
deduced in section 2 and what Sogami and Koizumi call a {\it periodic structure} in \cite{35}.\\
\indent Let $x=t+2\pi n$, where $t$ is confined to the \textit{partial interval} $t\in\left[-\pi,\pi\right]$ and substituting \eqref{5.4} in the right-hand side of \eqref{5.5}, we obtain after some straightforward calculations
\begin{eqnarray}\label{5.7}
\mathbf{1}_{\mathfrak{E}}
&=&\int_{-\pi}^{\pi}\rmd t\,
\overline{\mathbf{\Lambda}_s^{(\pm)}}(t)\sigma^{(\pm)}_s(t)\mathbf{\Lambda}_s^{(\pm)}(t), \nonumber \\
&=&\sum_{n=-\infty}^{+\infty}\int_{-\pi}^{+\pi}
\exp\left[-n\left(\gamma^{(\pm)}_s+\overline{\gamma^{(\pm)}_s}\right)
-2n\frac{\nu^{(\pm)}}{2}\frac{s^2}{\pi}\,t-2n^2\frac{\nu^{(\pm)}}{2}\,s^2\right]
\,\overline{\mathbf{\Lambda}_s^{(\pm)}}(t)\mathbf{\Lambda}_s^{(\pm)}(t)\,\rmd t,
\end{eqnarray}
and identifying with \eqref{5.5}, we can express the weight function $\sigma^{(\pm)}_s(t)$ by
\begin{eqnarray}\label{5.8}
\sigma^{(\pm)}_s(t)
&=&\sum_{n=-\infty}^{+\infty}\exp\left(-2\,n\Gamma^{(\pm)}-2\,n\frac{\nu^{(\pm)}s^2}{2\pi}\,t-n^2\nu^{(\pm)}\,s^2\right), \nonumber \\
&=&\sum_{n=-\infty}^{+\infty}\exp\left(2\,\rmi nz^{(\pm)}_s+\rmi n^2\pi\tau^{(\pm)}_s\right), \nonumber \\
&=&\vartheta_3\left(z^{(\pm)}_s(t)\Big|\tau^{(\pm)}(s)\right),
\end{eqnarray}
where the real parameter $\Gamma^{(\pm)}$ is given by
\begin{eqnarray}\label{5.9}
\Gamma^{(\pm)} &\equiv&
\frac{\gamma^{(\pm)}_s+\overline{\gamma^{(\pm)}_s}}{2}=1\pm 2\sqrt{\frac{\delta-1}{\delta(\delta+3)}}.
\end{eqnarray}
\indent As we can see, the weight function \eqref{5.8} is positive-definite and coincides exactly with the well-known elliptic Jacobi $\vartheta_3$-function \cite{36}, where its well-defined quasiperiodicities $z^{(\pm)}_s(t)$ and $\tau^{(\pm)}(s)$ depend linearly on the variable $t$ and the parameter $s$, respectively,
\begin{eqnarray}
z^{(\pm)}_s(t)  &=& \rmi\left(\Gamma^{(\pm)}_{\rm Re}+\nu^{(\pm)}\frac{s^2}{2\pi}\,t\right), \label{5.10} \\
\tau^{(\pm)}(s) &=& \rmi\frac{\nu^{(\pm)}s^2}{\pi}. \label{5.11}
\end{eqnarray}
\indent This series, \eqref{5.8}, converges for all finite $z^{(\pm)}_s(t)$ with the condition ${\rm Im}\left[\tau^{(\pm)}(s)\right]>0$, i.e. $\nu^{(\pm)}>0$, must be fulfilled. The latter condition, in conjunction with \eqref{5.9}, gives the following intervals according to the appropriate sign
\begin{eqnarray*}
  {\rm Positive\,\,sign} \,(+)&:& \delta\in\Delta^{(+)}=\left]-3,0\right[\cup\left]1,+\infty\right[, \\
  {\rm Negative\,\,sign} \,(-)&:& \delta\in\Delta^{(-)}=\left]-3,-(1+\sqrt{13})/2\right[\cup\left]1,(-1+\sqrt{13})/2\right[.
\end{eqnarray*}
\indent It is worth noting that the expression of $\sigma^{(\pm)}_s(t)$ obtained, \eqref{5.8}, is valid only for the weak-behavior expansion. However, the procedure of getting the expression of $\sigma^{(\pm)}_{\rm strong}(t)$ in the strong-behavior limit using \eqref{4.19} naturally introduces some difficulties of computation, which we do not resolve.

\section{Conclusion}%

\noindent In this paper, we have established a new expansion formulas in the cases of weak- and strong-behavior limits for the generalized $q$-deformed HO in order to solve the eigenvalue equation of BG CS. We started by applying a weak-behavior to our system up to the first order, and solving the Riccati differential equation, allowed us to deduce the weak expansions of the generalized $q$-deformed HO CS as well as their associated weak-eigenvalues. In the sequel, we have performed the resummation of the deduced results, using VPT, in order to describe our system in the strong-behavior limit. The corresponding strong $q$-deformed HO CS and their appropriate strong-eigenvalue expansions are obtained.\\
\indent Finally, we have shown that the simple and explicit form of the deduced $q$-deformed HO CS has enabled us to establish the relation of resolution of unity by means of a particular measure, the elliptic Jacobi $\vartheta_3$-function, on the unit circle which seems to be especially suited for this problem in order to bring out the nature of the space of functions.

\end{document}